\documentclass{optica-article}

\journal{opticajournal} 

\articletype{Research Article}

\usepackage{lineno}
\usepackage{caption}
\usepackage{subcaption}
\usepackage{float}
\usepackage[T1]{fontenc}

\begin{document}

\title{Photodiode quantum efficiency for 2-\textmu m light in the signal band of gravitational wave detectors} 

\author{Julian Gurs\authormark{1,\dag}, Nils S\"ultmann\authormark{1,\dag}, Christian Darsow-Fromm\authormark{2}, Sebastian Steinlechner\authormark{3,4} and Roman Schnabel\authormark{1,*}}

\address{\authormark{1}Institut f\"ur Quantenphysik \& Zentrum f\"ur Optische Quantentechnologien, Universit\"at Hamburg, Luruper Chaussee 149,22761 Hamburg, Germany\\
\authormark{2}Institut f\"ur Experimentalphysik, Universit\"at Hamburg, Luruper Chaussee 149, 22761 Hamburg, Germany\\
\authormark{3}Faculty of Science and Engineering, Maastricht University, Duboisdomein 30, 6229 GT Maastricht, The Netherlands\\
\authormark{4}Nikhef, Science Park 105, 1098 XG Amsterdam, The Netherlands\\
\authormark{\dag}These authors contributed equally to this work; the order is determined alphabetically.\\
\authormark{*}Corresponding author:
\url{roman.schnabel@uni-hamburg.de} (R. Schnabel)}

\begin{abstract*}
Quantum technologies with quantum correlated light require photodiodes with near-perfect `true' quantum efficiency, the definition of which adequately accounts for the photodiode dark noise. Future squeezed-light-enhanced gravitational wave detectors could in principle achieve higher sensitivities with a longer laser wavelength around 2\,\textmu m. Photodiodes made of extended InGaAs are available for this range, but the true quantum efficiency at room temperature and the low frequency band of gravitational waves is strongly reduced by dark noise. Here we characterize the change in performance of a commercial extended-InGaAs photodiode versus temperature. While the dark noise decreases as expected with decreasing temperature, the detection efficiency unfortunately also decreases monotonically. Our results indicate the need for a dedicated {new} design of photodiodes for gravitational wave detectors using 2-\textmu m laser light.
\end{abstract*}

\section{Introduction}
The first detections of gravitational waves (GWs) since 2015 by LIGO and Virgo \cite{GW150914,GW170814} have accelerated the development work for new generations of GW observatories such as the Einstein Telescope \cite{Punturo2010}, LIGO Voyager \cite{Adhikari2020} and Cosmic Explorer \cite{Reitze2019}.
A fundamental source of noise in GW observatories is the thermally excited motion of the mirror surfaces. It leads to so-called thermal noise in the observatories' output light with contributions of various mechanisms. 
In future, the thermal noise is to be further reduced, also by cryogenic cooling of the mirrors, as is already the case at KAGRA \cite{KAGRA2019}.
Cryogenically cooled mirrors made of crystalline silicon with mirror coatings consisting of alternating layers of amorphous silicon and silicon nitride are very promising in this respect \cite{SteinlechnerJ2018}. A side effect would be that instead of the current wavelength of 1064\,nm, a wavelength about twice as long at around 2\,\textmu m would have to be used.
{A longer wavelength also has the advantage of reducing scattering caused by optical surface defects.}
GW observatories use quantum-correlated (squeezed) laser light \cite{Schnabel2010,Schnabel2017,LSC2011,Grote2013,Tse2019,Acernese2019}, the advantage of which can only be utilized with photodiodes having very high detection efficiency combined with low dark noise \cite{Vahlbruch2016}. 

Here we present the measurement of changes in quantum efficiency of a commercial extended-InGaAs photodiode when cooled from room temperature down to 4 K. The transimpedance amplifier and other electronic components were kept at room temperature. Our results show the expected decrease in photodiode dark noise. 
However, it turned out that the detection efficiency of the photodiode also decreased with decreasing temperature. This shows that cooling of current photodiodes alone is not sufficient to achieve ultra-high true quantum efficiency for 2\,\textmu m laser light.

The definition of the detection efficiency $\eta_{\rm DE}$ does not take into account photodiode dark noise. 
In the field of quantum technologies, which includes quantum sensing, quantum communication and quantum computing, electrons that are elevated into the conduction band with thermal energy are just as undesirable as undetected photons.
We define the quantum efficiency $\eta_{\rm QE}$, which takes photodiode dark noise into account. The following equations compare the two definitions. 
\begin{equation}
\label{eq:1}
\eta_{\rm DE} = \frac{U_{\rm tot} - U_{\rm dark}}{U_{\rm perf}}, \hspace{16mm}
\eta_{\rm QE} = \frac{U_{\rm tot} - U_{\rm dark}}{U_{\rm perf} + U_{\rm dark}} \; ,
\end{equation}
where the voltages $U$ are transimpedance amplified photo currents measured on continuous-wave light. $U_{\rm tot}$ is the total measurement voltage, $U_{\rm dark}$ is the contribution with laser light switched off. $U_{\rm perf}$ is the perfect case, i.e.~the usually unknown voltage that the measurement system would provide if every photon of the light beam (that is focussed on the photodiode surface) would be transferred into exactly one photo electron, combined with zero dark current. 
For the manufacturer of photodiodes, the detection efficiency $\eta_{\rm DE}$ is the relevant parameter because the problem of dark noise also depends on the light power used and the noise of the transimpedance amplifier. For the user of quantum technologies, on the other hand, $\eta_{\rm QE}$, the quantum efficiency (of the entire measuring apparatus), is the relevant parameter. As an example, the quantum efficiency is correctly only 50\% if an equally large dark current adds to that from perfect detection efficiency. In the literature, the term quantum efficiency is often used as a synonym for detection efficiency.

We carried out all measurements with continuous-wave laser light with 2128\,nm wavelength.
This particular wavelength is a promising candidate for future GW detectors, as it can be produced by degenerate optical parametric oscillation (DOPO) from the existing ultra-stable laser light at 1064 nm. Corresponding evidence has already been provided \cite{Darsow-Fromm2020,Gurs2024,Gurs2025}. Squeezed light can be produced at this wavelength \cite{Darsow-Fromm2021}. Other wavelengths in the 2\,\textmu m range have also been researched \cite{Tang2012,Mansell2018,Yap2019}.

\section{Experimental setup}
\begin{figure}[H]
    \captionsetup[subfigure]{justification=centering}
    \centering
    \begin{subfigure}[b]{0.7\textwidth}
        \centering
        \includegraphics[width=\linewidth]{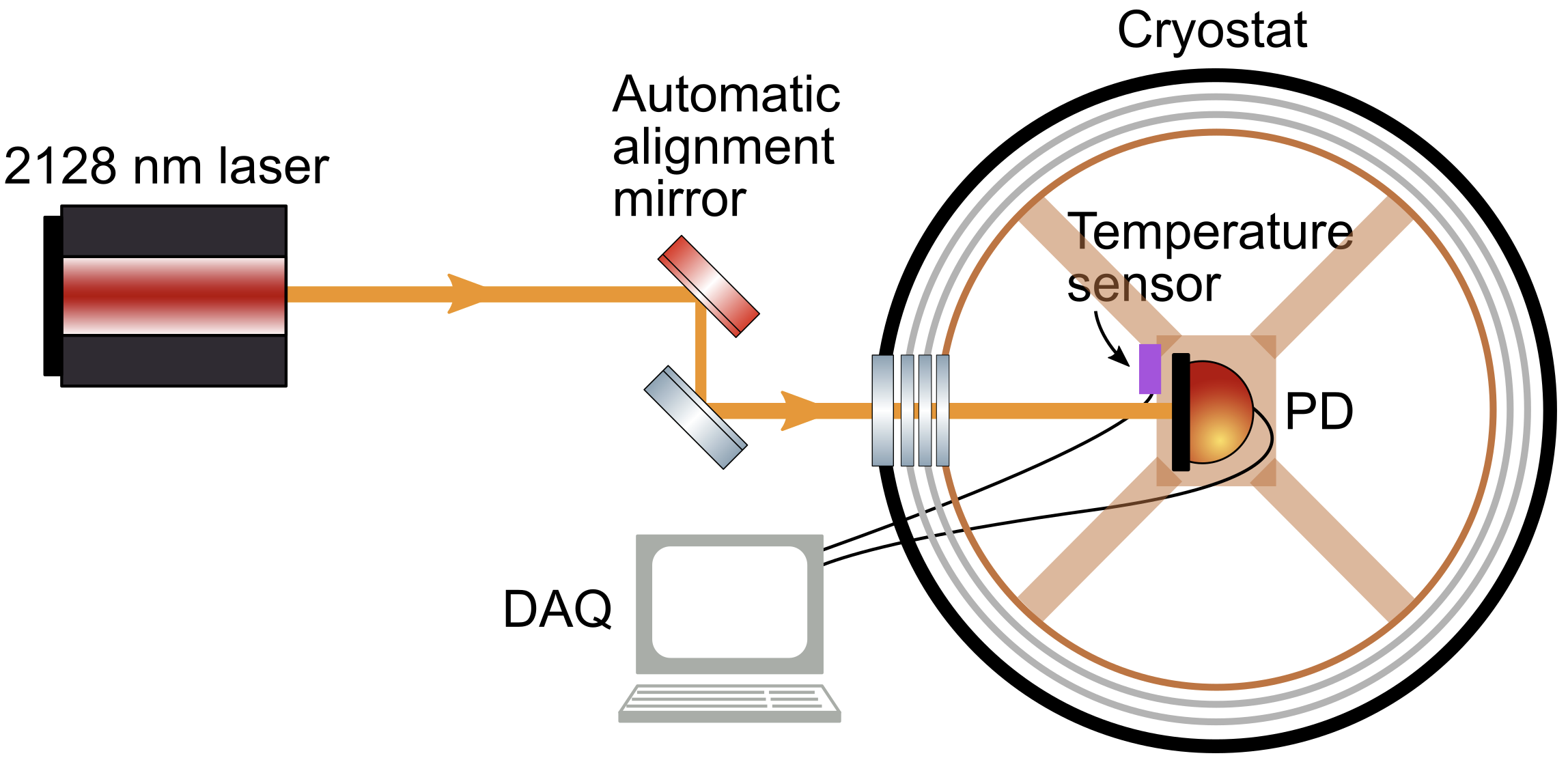}
    \end{subfigure}\\
    \hfill \\
    \begin{subfigure}[b]{0.65\textwidth}
        \centering
        \includegraphics[width=\linewidth]{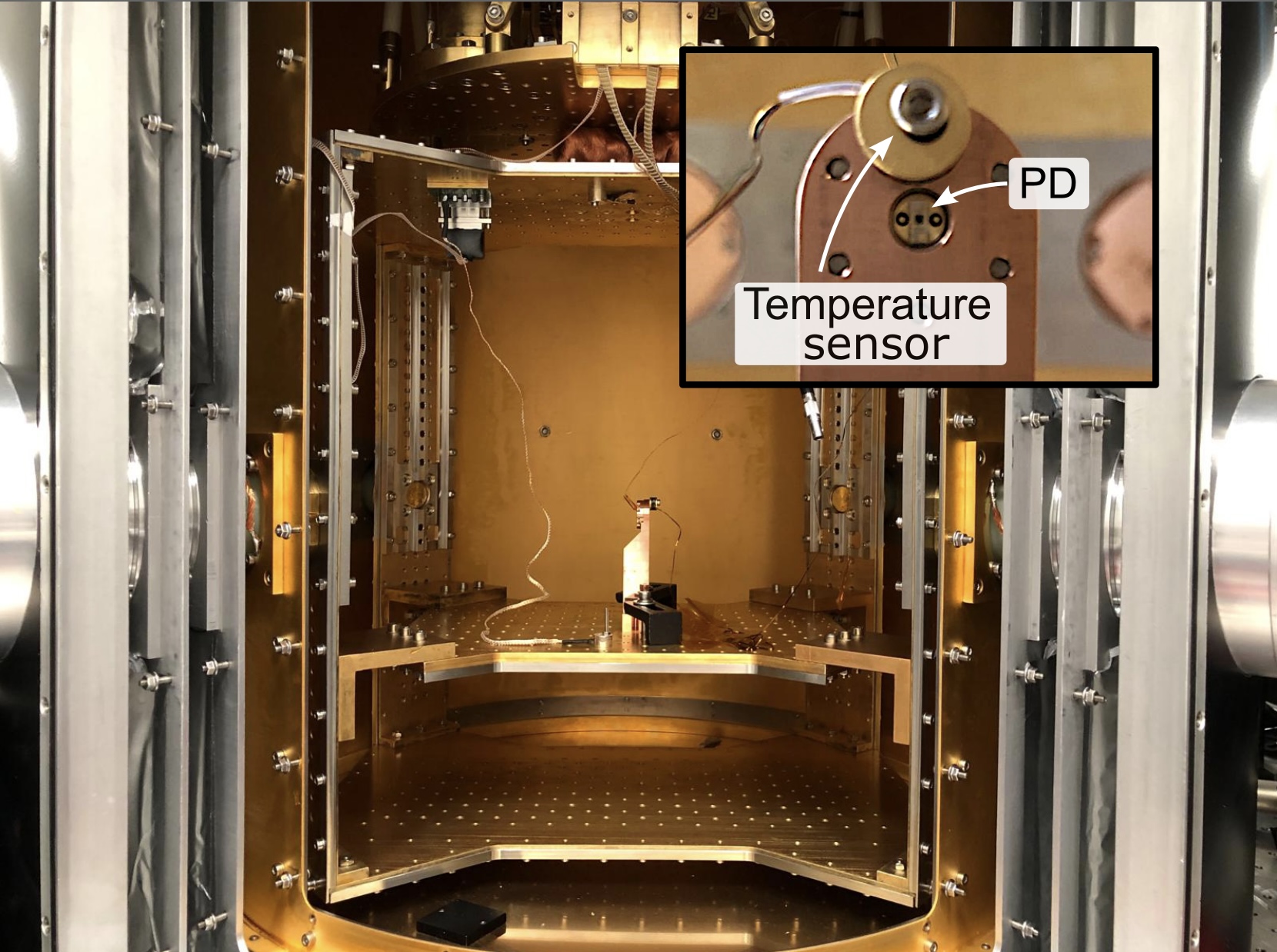}
    \end{subfigure}
    \caption{A \textsc{Thorlabs} FD05D photodiode (PD) was placed in a helium cryostat and cooled down to 4\,K  
    {within 48\,h.} 
    The photo current was converted to a voltage by a high-gain self-built transimpedance amplifier located in room temperature environment. During cool-down and warm-up, the laser light on the photodiode was switched on and off every 5 minutes and the photo voltage values were continuously sampled and stored at a rate of 1\,kHz. This measurement data provided dark current, total photo current and their spectral densities. In parallel, the temperature of the photodiode was measured with a sensor in a common copper block. The stability of the light power on the photodiode was ensured by automatic beam centering and checked by a separate measurement on the input beam. DAQ: data acquisition system.
  }
    \label{fig1:setup}
\end{figure}

Figure\,\ref{fig1:setup} shows the schematic of our experiment and a photograph of the photodiode inside our cryostat. Continuous-wave laser light of 0.1\,mW at 2128\,nm from a degenerate optical parametric oscillator \cite{Darsow-Fromm2020} was directed onto the surface of the \textsc{Thorlabs} FD05D extended InGaAs photodiode (designed for the wavelength range of 0.9 to 2.6\,\textmu m; diameter of the active area: 0.5\,mm. 
{We removed the photodiode window before taking the measurements to avoid unnecessary photon losses (of approx. 3\%). The remaining reflectivity was that at the semiconductor surface. We determined it to be approximately 0.8\%.}
The incident light power was controlled and stable over the days when the measurements were performed.
The cryostat was an \textsc{Entropy} closed-cycle helium cryostat with free-beam optical view ports 
{with a clear aperture of 10\,mm diameter.}
Three heat shield layers with anti-reflective coated windows provided shielding from thermal radiation. 
The cold experimentation chamber was mechanically decoupled from the vibrating cooler unit and mechanically connected to the optical table. 
One of the steering mirrors was actively controlled to keep the laser spot in the centre of the photodiode compensating movement due to thermal shrinking and expansions during cool-down and warm-up.
{We used a gradient descent with momentum algorithm \cite{Qian1999}, which steered every five minutes while not taking data the mirror for maximum power on the PD. This was achieved by iteratively checking the change in PD response and then moving the mirror based on the improvement in response to earlier measurements.}
The photodiode was mounted in a copper block with good thermal contact, as seen in the zoom in the photograph in Fig.\,\ref{fig1:setup}. A cernox temperature sensor from \textsc{Lakeshore} was installed in the immediate vicinity. The self-built transimpedance amplifier (gain of 300\,V/mA) of the photodiode was housed outside of the cryostat. Due to its high gain, the photodiode dark noise was significantly higher than other non-optical noise in our measurements. 
The optical signal of our detection efficiency measurements was broad-band optical power noise of the incident 2128\,nm light of 0.1\,mW, see Fig.\,\ref{fig2:DarkNoise}\,(left). The rather low power minimized heating of the photodiode. 

\section{Results}
\begin{figure}[H]
    \captionsetup[subfigure]{justification=centering}
    \centering
    \begin{subfigure}[b]{0.493\textwidth}
        \centering
        \includegraphics[width=\linewidth]{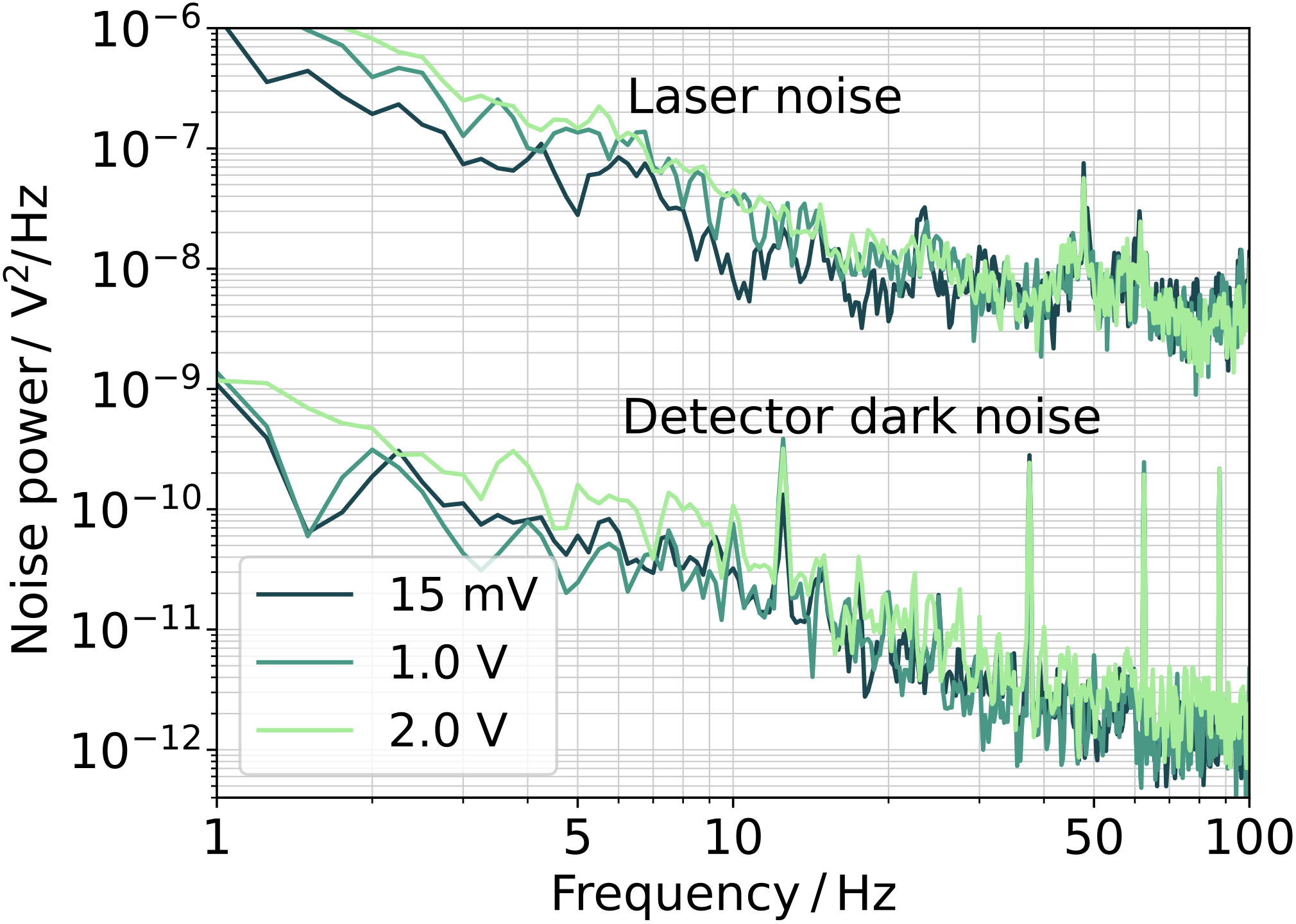}
    \end{subfigure}
    \begin{subfigure}[b]{0.50\textwidth}
        \centering
        \includegraphics[width=\linewidth]{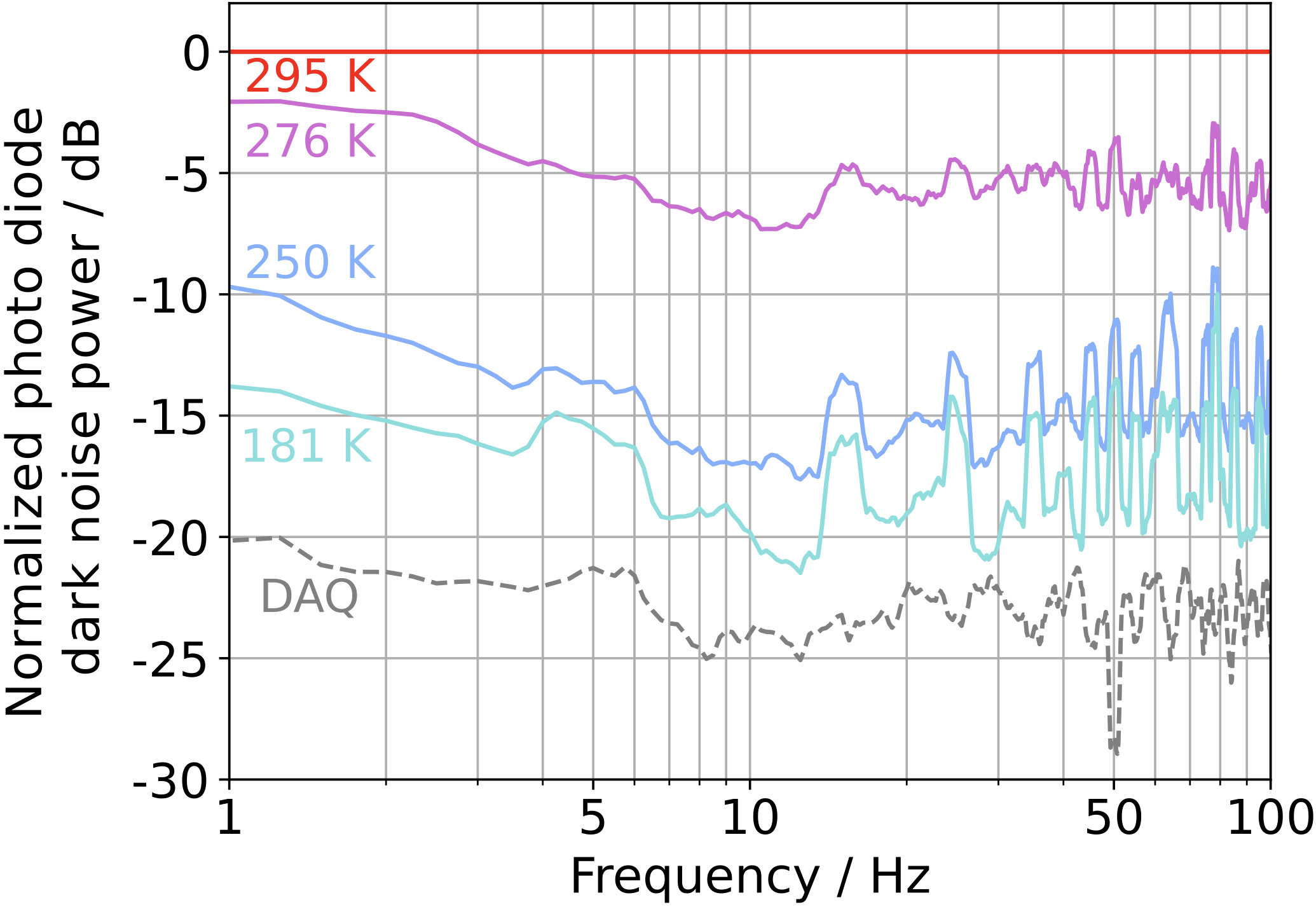}
    \end{subfigure}
    \caption{
    Noise power spectral densities from 1 to 100\,Hz.
    Left: Comparison of laser power noise from 0.1\,mW at 2128\,nm and dark noise measured with the photodiode FD05D at 144\,K for three bias voltages (The resolution bandwidth, RBW: 0.1\,Hz, no averaging)
    Right: Photodiode dark-noise at four temperatures, normalized to that at room temperature. With decreasing temperature, detector dark noise strongly decreased as expected. The photodiode bias voltage was 0.39\,V. (RBW: 0.25\,Hz, 50 times averaged).
    }
    \label{fig2:DarkNoise}
\end{figure}

Figure \ref{fig2:DarkNoise}\,(left) provides an overview of the spectral noise powers in our experiment for sideband frequencies that are relevant for future earth-based cooled GW detectors. The photodiode temperature here is exemplary at 144\,K. The power noise of 0.1\,mW of the 2128\,nm light (our `signal') is significantly higher than the dark noise of the FD05D photodiode.\\
Figure \ref{fig2:DarkNoise}\,(right) shows measured dark noise power spectral densities of the photodiode at room temperature and three different lower temperatures. The lowest noise powers were already achieved at around 180\,K with a reduction of more than 15\,dB. At further reduced temperatures, the dark noise power did not drop any further, which was probably due to the dark noise of the room-temperature transimpedance amplifier stage. (The noise power of our data acquisition system (dashed) was approximately 5\,dB below the lowest dark noise of the measurement system.)\\
The general reduction in dark noise with lower temperature is expected and well understood, as dark noise is essentially noise due to thermal energy. The thermal energy of the charge carriers leads to a statistical probability of overcoming the band gap between the valence band and the conduction band even without light incidence.
Our dark noise spectra show several peaks, which we were able to attribute to a 50\,Hz ground loop of the mains and the temperature data recording at a rate of 5\,Hz.

\begin{figure}[H]
     \centering
     \includegraphics[width=0.7\linewidth]{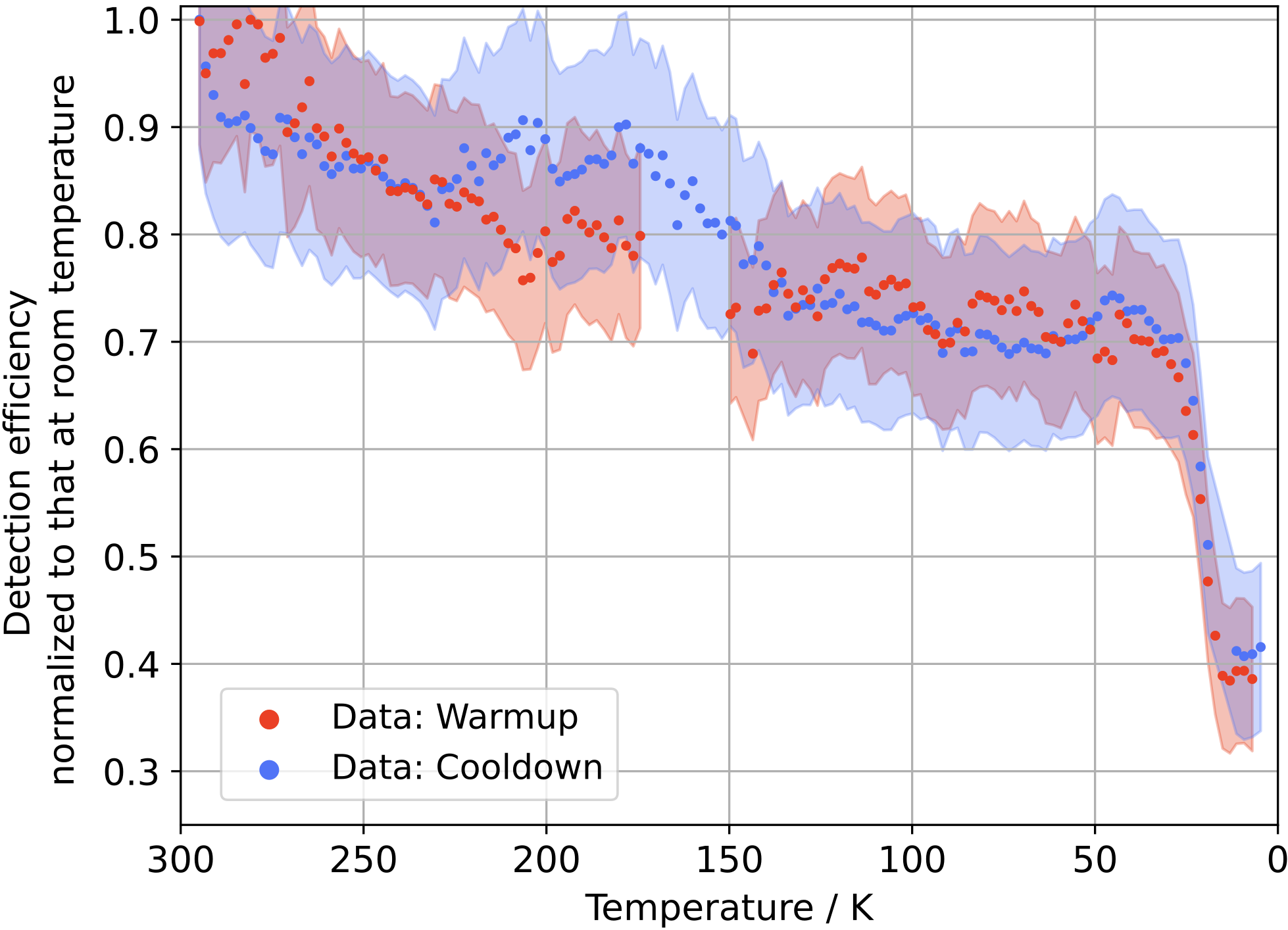}
     \caption{Change of {the detection efficiency} of PD FD05D with temperature, normalized to that at room temperature. The data was collected continuously over one full cooling cycle. Blue points were taken while cooling down over a time frame of 48 hours. Red points show the {natural} warming up over 16 days {(cooling switched off, no heating)}. The missing data in the warm-up is due to a computer failure. A small hysteresis of the data was observed. The shaded areas indicate our estimation of systematic error bars due to fluctuations in the light power. All data was measured at a photodiode bias voltage of 0.39\,V.
     }
     \label{fig3:DE}
 \end{figure}

{
By cooling our extended InGaAs photodiode to temperatures below 200 K, we achieved a level of dark noise that was irrelevant when measuring laser power noise in the frequency band from 1 to 100 Hz (Fig.\,\ref{fig2:DarkNoise}). The detection efficiency largely approximated the quantum efficiency according to Eq.\,(\ref{eq:1}).
However, the optimization of highly efficient photodiodes via cooling is only successful if the detection efficiency does not decrease as a result of cooling.  
The measurement results of this investigation on our FD05D photodiode are shown in Fig.\,\ref{fig3:DE}.
The data was recorded over a cool-down period of 48 hours and a warm-up period of about 16 days. 
Measured was 0.1\,mW stable laser light at 2128\,nm with the noise power shown in Fig.\,\ref{fig2:DarkNoise}\,(left). 
The graph shows the {\it normalized} detection efficiency versus temperature. It was normalized to that at room temperature, i.e.
\begin{equation}
\label{eq:2}
\eta_{\rm DE, norm}(T) 
= \frac{\eta_{\rm DE}(T)}{\eta_{\rm DE}({\rm 293\,K})} 
= \frac{U_{\rm tot}(T) - U_{\rm dark}(T)} {U_{\rm tot}({\rm 293\,K}) - U_{\rm dark}({\rm 293\,K})} \; .
\end{equation}
The voltages ($U$) in Eq.\,({\ref{eq:2}}) were measured every five minutes. The photo voltage was sampled for four seconds at {a rate of} 10\,MHz when the laser light was either switched on ($U_{\rm tot}$) or off ($U_{\rm dark}$) and averaged to a single DC mean value. Values within the same 1-K temperature intervals were further combined to a unified DC mean value.
}

{Fig.\,\ref{fig3:DE} shows a continuously decreasing detection efficiency when cooling down from room temperature, which is an unfortunate result. At 250\,K (where the dark noise drops to below 10\%, see Fig.\,\ref{fig2:DarkNoise} on the right) the detection efficiency is reduced by 15\%. Photodiodes for future gravitational wave detectors must have a detection efficiency (and quantum efficiency) of approximately 99\% or better. A deterioration of 15\% due to cooling would be unacceptable. A decrease in detection efficiency with decreasing temperature is not unknown and was recently also observed in Ref.\cite{Bajpai2022} (Fig.\,3) and Ref.\cite{Kuhlbusch2024} (Fig.\,6). 
Below 25\,K, our detection efficiency further dropped sharply and reached a value of approximately 40\% at 10\,K compared to that at room temperature. A sharp efficiency drop below a characteristic temperature is well-known and explained as the result of an enlarged band gap \cite{Gaskill1990,Rogalski2009,Kuhlbusch2024}. In principle, the viewport system of our cryostat could have led to a falsely exaggerated decrease in detection efficiency. However, the cryostat is specially designed to avoid this. The temperature-related shrinkage of the materials is compensated for in such a way that the free aperture does not decrease when the temperature changes. We checked this during the measurement and were unable to detect any scattered laser light at the apertures.}

\section{Conclusion}
Quantum technologies require photodiodes with true quantum efficiencies close to one (> 99\%), i.e. virtually perfect detection efficiencies with simultaneously negligible dark noise. For GW detectors, these properties must persist in the low-frequency signal band of the targeted GWs. For future terrestrial GW detectors with cryogenically-cooled mirrors, this applies to the sideband of approx.~1 to 100 Hz.

{With this work, we have investigated the change of detection efficiency of a commercial extended-InGaAs photodiode for 2128\,nm light and the change of dark noise at low sideband frequencies versus temperature. The photodiode was designed for high detection efficiency at room temperature for the wavelength range from 0.9 to
2.6\,\textmu m. With a hand-selected pair of the same photodiode model, we were able in the past to demonstrate a squeeze factor of 7\,dB at a sideband frequency of 2\,MHz, from which we concluded that the detection efficiency was approximately 93\% at 2128\,nm at room temperature \cite{Darsow-Fromm2021}.
}

{The positive result of our measurements is that a temperature of approximately {200\,K} already reduces the photodiode dark noise in the mentioned low frequency band by more than 15\,dB (Fig.\,\ref{fig2:DarkNoise}). 
Another positive result is that the photodiode showed no degradation of performance from being {cooled down to 5\,K four times and} operated at bias voltages significantly higher then specified by the manufacturer.}

{The negative result of our measurement is that with cooling {down to 200\,K, the detection} efficiency of the same photodiode decreased by ($15 \pm 5$)\% compared to that at room temperature. 
A concept for highly efficient photodiodes that requires cooling to reduce dark noise but loses detection efficiency at lower temperatures is not suitable for quantum technologies.
For future cryogenic 2\,\textmu m\,-\,GW detectors targeting the scientifically interesting frequency range from 1 to 100\,Hz,
a new type of photodiode must be developed. The design must ensure quantum efficiency of more than 99\% at these low frequencies with an active area diameter of approximately 3\,mm, which is a typical size in GW detection.} 

\section{Back matter}
 \begin{backmatter}


\bmsection{Acknowledgment}
This research has been funded by the Germany Federal Ministry of Education and Research, grant no. 05A20GU5. The cryostat was financed by the Deutsche Forschungsgemeinschaft and the state of Hamburg, grant no INST 187/428-1 FUGG. S.~Steinlechner acknowledges funding by the Province of Limburg and the Dutch Research Council (NWO), grant VI.Vidi.213.127. This article has LIGO document number P2500292.


\bmsection{Data Availability Statement}
Data underlying the results presented in this paper are not publicly available at this time but may be obtained from the authors upon reasonable request.

\end{backmatter}


\begin{thebibliography}{10}
\newcommand{\enquote}[1]{``#1''}

\bibitem{GW150914}
B.~P.~Abbott~\emph{et al.}, \enquote{{Observation of Gravitational Waves from a
  Binary Black Hole Merger},} {\protect\JournalTitle{Physical Review Letters}}
  \textbf{116}, 061102 (2016).

\bibitem{GW170814}
B.~P.~Abbott~{\it et al.}, \enquote{{GW170814: A Three-Detector Observation of
  Gravitational Waves from a Binary Black Hole Coalescence},}
  {\protect\JournalTitle{Physical Review Letters}} \textbf{119}, 141101 (2017).

\bibitem{Punturo2010}
M.~Punturo~{\it et al.}, \enquote{{The third generation of gravitational wave
  observatories and their science reach},} {\protect\JournalTitle{Classical and
  Quantum Gravity}} \textbf{27}, 084007 (2010).

\bibitem{Adhikari2020}
R.~X. Adhikari~\emph{et al.}, \enquote{{A cryogenic silicon interferometer for
  gravitational-wave detection},} {\protect\JournalTitle{Classical and Quantum
  Gravity}} \textbf{37}, 165003 (2020).

\bibitem{Reitze2019}
D.~Reitze, R.~X. Adhikari, S.~Ballmer, \emph{et~al.}, \enquote{Cosmic
  {{Explorer}}: {{The U}}.{{S}}. {{Contribution}} to {{Gravitational}}-{{Wave
  Astronomy}} beyond {{LIGO}},} {\protect\JournalTitle{arXiv:1907.04833
  [astro-ph, physics:gr-qc]}}  (2019).

\bibitem{KAGRA2019}
T.~Akutsu~{\it et al.}, \enquote{{KAGRA: 2.5 generation interferometric
  gravitational wave detector},} {\protect\JournalTitle{Nature Astronomy}}
  \textbf{3}, 35--40 (2019).
  
  \bibitem{SteinlechnerJ2018}
J.~Steinlechner, I.~W. Martin, A.~S. Bell, \emph{et~al.},
  \enquote{{Silicon-Based Optical Mirror Coatings for Ultrahigh Precision
  Metrology and Sensing},} {\protect\JournalTitle{Physical Review Letters}}
  \textbf{120}, 263602 (2018).

\bibitem{Schnabel2010}
R.~Schnabel, N.~Mavalvala, D.~E. McClelland, and P.~K. Lam, \enquote{{Quantum
  metrology for gravitational wave astronomy.}} {\protect\JournalTitle{Nature
  communications}} \textbf{1}, 121 (2010).

\bibitem{Schnabel2017}
R.~Schnabel, \enquote{{Squeezed states of light and their applications in laser
  interferometers},} {\protect\JournalTitle{Physics Reports}} \textbf{684},
  1--51 (2017).

\bibitem{LSC2011}
J.~Abadie~\emph{et al.}, \enquote{{A gravitational wave observatory operating
  beyond the quantum shot-noise limit},} {\protect\JournalTitle{Nature
  Physics}} \textbf{7}, 962--965 (2011).

\bibitem{Grote2013}
H.~Grote, K.~Danzmann, K.~L. Dooley, \emph{et~al.}, \enquote{{First Long-Term
  Application of Squeezed States of Light in a Gravitational-Wave
  Observatory},} {\protect\JournalTitle{Physical Review Letters}} \textbf{110},
  181101 (2013).

\bibitem{Tse2019}
M.~Tse~\emph{et al.}, \enquote{{Quantum-Enhanced Advanced LIGO Detectors in the
  Era of Gravitational-Wave Astronomy},} {\protect\JournalTitle{Physical Review
  Letters}} \textbf{123}, 231107 (2019).

\bibitem{Acernese2019}
F.~Acernese~\emph{et al.}, \enquote{{Increasing the Astrophysical Reach of the
  Advanced Virgo Detector via the Application of Squeezed Vacuum States of
  Light},} {\protect\JournalTitle{Physical Review Letters}} \textbf{123},
  231108 (2019).

\bibitem{Vahlbruch2016}
H.~Vahlbruch, M.~Mehmet, K.~Danzmann, and R.~Schnabel, \enquote{{Detection of
  15 dB Squeezed States of Light and their Application for the Absolute
  Calibration of Photoelectric Quantum Efficiency},}
  {\protect\JournalTitle{Physical Review Letters}} \textbf{117}, 110801 (2016).

\bibitem{Darsow-Fromm2020}
C.~Darsow-Fromm, M.~Schr{\"{o}}der, J.~Gurs, \emph{et~al.},
  \enquote{{Highly-efficient generation of coherent light at 2128 nm via
  degenerate optical-parametric oscillation},} {\protect\JournalTitle{Optics
  Letters}} \textbf{45}, 6194--6197 (2020).

\bibitem{Gurs2024}
J.~Gurs, N.~Bode, C.~Darsow-Fromm, \emph{et~al.}, \enquote{{Conversion of 30 W
  laser light at 1064 nm to 20 W at 2128 nm and comparison of relative power
  noise},} {\protect\JournalTitle{Classical and Quantum Gravity}} \textbf{41},
  245008 (2024).

\bibitem{Gurs2025}
J.~Gurs, M.~Korobko, C.~Darsow-Fromm, \emph{et~al.}, \enquote{{Coherent noise
  suppression at high-efficiency wavelength doubling for high-precision
  experiments},} {\protect\JournalTitle{Optics and Laser Technology}}
  \textbf{183} (2025).

\bibitem{Darsow-Fromm2021}
C.~Darsow-Fromm, J.~Gurs, R.~Schnabel, and S.~Steinlechner, \enquote{{Squeezed
  light at 2128 nm for future gravitational-wave observatories},}
  {\protect\JournalTitle{Optics Letters}} \textbf{46}, 5850 (2021).

\bibitem{Tang2012}
Y.~Tang, C.~Huang, S.~Wang, \emph{et~al.}, \enquote{High-power narrow-bandwidth
  thulium fiber laser with an all-fiber cavity,} {\protect\JournalTitle{Opt.
  Express}} \textbf{20}, 17539--17544 (2012).

\bibitem{Mansell2018}
G.~L. Mansell, T.~G. McRae, P.~A. Altin, \emph{et~al.}, \enquote{Observation of
  {{Squeezed Light}} in the 2\,\textmu m {{Region}},}
  {\protect\JournalTitle{Physical Review Letters}} \textbf{120} (2018).

\bibitem{Yap2019}
M.~J. Yap, D.~W. Gould, T.~G. McRae, \emph{et~al.}, \enquote{Squeezed vacuum
  phase control at 2\,\textmu m,} {\protect\JournalTitle{Optics Letters}}
  \textbf{44}, 5386--5389 (2019).

\bibitem{Qian1999}
N.~Qian, \enquote{On the momentum term in gradient descent learning
  algorithms,} {\protect\JournalTitle{Neural networks}} \textbf{12}, 145--151
  (1999).

\bibitem{Bajpai2022}
R. Bajpai, T.~Tomaru, K.~Yamamoto, \emph{et~al.}, \enquote{A laser interferometer accelerometer for vibration sensitive cryogenic experiments} {\protect\JournalTitle{Measurement Science and Technology}}
  \textbf{33}, 085902 (2022).
  
  \bibitem{Kuhlbusch2024}
T.~J. Kuhlbusch, M.~Zeoli, R.~Joppe, \emph{et~al.}, \enquote{Characterizing
  1550 nm optical components down to 8 K,} {\protect\JournalTitle{Cryogenics}}
  \textbf{142}, 103895 (2024).
  
  \bibitem{Gaskill1990}
D.~K. Gaskill, N.~Bottka, L.~Aina, and M.~Mattingly, \enquote{{Band-gap
  determination by photoreflectance of InGaAs and InAlAs lattice matched to
  InP},} {\protect\JournalTitle{Applied Physics Letters}} \textbf{56},
  1269--1271 (1990).

\bibitem{Rogalski2009}
A.~Rogalski, J.~Antoszewski, and L.~Faraone, \enquote{Third-generation infrared
  photodetector arrays,} {\protect\JournalTitle{Journal of Applied Physics}}
  \textbf{105}, 091101 (2009).

\end{thebibliography}

\end{document}